# Trajectory Tracking for MmWave Communication Systems via Cooperative Passive Sensing

Chao YU, Bojie Lv, Haoyu QIU, Rui WANG

(Department of Electronic and Electrical Engineering,
Southern University of Science and Technology (SUSTech), Shenzhen 518000, China)

**Abstract:**
In this paper, a cooperative passive sensing framework for millimeter wave (mmWave) communication systems is proposed and demonstrated in a scenario with one mobile signal blocker. Specifically, in the uplink communication with at least two transmitters, a cooperative detection method is proposed for the receiver to track the blocker's trajectory, localize the transmitters and detect the potential link blockage jointly. To facilitate detection, the receiver collects the signal of each transmitter along a line-of-sight (LoS) path and a non-line-of-sight (NLoS) path separately via two narrow-beam phased arrays. The latter path should scatter at the mobile blocker, and hence it can be identified by the Doppler frequency. Comparing the received signals of both paths, the Doppler frequency and angle-of-arrival (AoA) of the NLoS path can be estimated. To resolve the blocker's trajectory and the transmitters' locations, the receiver should continuously track the mobile blocker to accumulate sufficient numbers of the Doppler frequency and AoA versus time observations. Finally, a gradient-descent-based algorithm is proposed for joint detection. With the reconstructed trajectory, the potential link blockage can be predicted. It is demonstrated that the system can achieve decimeter-level localization and trajectory estimation, and predict the blockage time with an error less than 0.1 s.

## 1 Introduction

Due to the short wavelength, millimeter-wave (mmWave) wireless communication is fragile to link blockage[1–3]. Fortunately, with the development of wireless sensing techniques, it is feasible to design a robust mmWave communication system, which can predict the link blockage, detect backup signal propagation paths, and mitigate the impact of signal-to-noise ratio (SNR) loss via predictive scheduling.

There have been a number of research efforts on the blockage prediction via out-of-band sensors[4–6] or in-band channel information[7–11]. For example, a camera was proposed to sense the communication environment and predict the mmWave link blockage via deep learning algorithms in Refs. [4–5]. A light detection and ranging (LiDAR) assisted proactive blockage prediction scheme was proposed in Ref. [6]. The use of out-of-band sensors not only increases the cost of communication systems, but also raises privacy issues, especially in indoor communication scenarios. With the in-band channel information, the variation of received signal strength (RSS) was proposed to predict blockage in Refs. [7, 11], where deep learning algorithms were developed to track the RSS variation patterns right before blockage. It was found in Ref. [8] that the diffraction effects of mmWave signals could be exploited in blockage prediction. Moreover, a protective beam was proposed in Ref. [9] to monitor the Doppler effect in the communication environment, so that potential link blockage can be forecasted. However, none of the above in-band sensing methods can track the trajectory of mobile blockers. Without the trajectory, these methods may not provide sufficient warning time before blockage, or they may lead to large false alarm probability. In our preliminary study[12], an mmWave blockage prediction method via passive sensing architecture was proposed, which could predict 90% of the line-of-sight LoS blockage with a sensing time of 1.4 s. It assumed location knowledge of the transmitter and receiver, as well as the constant velocity of the mobile blocker.

In fact, passive sensing is a promising approach to facilitating simultaneous sensing and data communication with half-duplexing transceivers[13]. By comparing the received signals of reference and surveillance channels, multi-antenna passive sensing techniques can detect the direction, distance and the raised Doppler frequency of a mobile target. It has been used for localization or trajectory tracking via WiFi signals[14] or long-term evolution (LTE) downlink signals[15]. To further improve the tracking accuracy, multiple signal transmitters or sensing receivers can be adopted to detect the Doppler frequency in different dimensions. For example, a handwriting tracking method via cooperative passive sensing of two receivers was proposed in Ref. [16], where the accuracy of handwriting reconstruction was at the millimeter-level. However, all these works assumed knowledge on the locations of signal transmitters and sensing receivers, which might not be easily obtained in mobile communication systems, especially in indoor scenarios.

In this paper, we would like to address the above issue by proposing a cooperative passive sensing method for joint trajectory tracking and device localization. Specifically, at least two uplink transmitters simultaneously transmit uplink signals to one receiver in different frequency bands, and there is one mobile blocker in the communication environment. The receiver adopts two narrow beams for receiving the uplink signal of each transmitter. One beam is aligned with the transmitter directly, and the other is aimed at the mobile blocker. Thus, the signal of LoS path is received by the former beam, and the scattered signal from the mobile blocker, which is with non-zero Doppler frequencies, is received by the latter one. By comparing the signals of the above two beams, the Doppler frequencies and angles-of-arrival (AoAs) of the mobile blocker can be continuously observed. Accumulating the above observations from two transmitters, we find the transmitters' locations and the blocker's trajectory can be jointly detected via a proposed gradient-descent-based algorithm. It is demonstrated that the system can achieve decimeter-level localization and trajectory estimation, and predict the blockage time with an error less than 0.1 s.

The remainder of this paper is organized as follows. In Section 2, an overview of the system is provided. In Section 3, the signal processing for passive sensing is introduced. Section 4 introduces the algorithms for joint trajectory tracking and transmitter localization, followed by the method of blockage prediction. The experiment results and analysis are provided in Section 5. Finally, conclusions are drawn in Section 6.

## 2 System Overview

In this paper, a trajectory tracking and blockage prediction method is proposed for mmWave uplink communications. Without a priori location knowledge of the transmitters and receiver, the proposed method can simultaneously track the trajectory of a mobile signal blocker and detect the locations of the transmitters with respect to the receiver. As a result, the potential blockage of the LoS path can be predicted. The overall system architecture is illustrated in Fig. 1. The proposed system consists of one mmWave receiver and at least two transmitters, where the receiver receives signals from all transmitters simultaneously by frequency division. The two transmitters are referred to as Transmitters 1 and 2, and their orthogonal communication bands are referred to as Bands 1 and 2, respectively. The transmitters and receiver can be the user equipment (UE) and base station (BS) of uplink communications, respectively.

Each mmWave transmitter delivers an information-bearing signal via a transmission beam. The signal arrives at the receiver via both the LoS path and the non-line-of-sight (NLoS) path scattered at the mobile blocker, as illustrated in Fig. 1. The receiver has as least two RF chains, each with a phased array. One of the phased array generates a receiving beam, namely a surveillance beam, to capture the NLoS signal in both frequency bands. Due to the mobility of the blocker, this receiving beam should periodically sweep the surrounding region. The other phased array generates two receiving beams towards the two transmitters, respectively. They could capture the LoS signals of the two transmitters in two frequency bands with s high signal-to-noise ratio (SNR), respectively. The LoS paths from the two transmitters are referred to as the reference channels 1 and 2, and their receiving beams are referred to as the reference beams 1 and 2, respectively. Additionally, the signal propagation paths from both transmitters and scattered by the mobile blocker are called the surveillance channels 1 and 2, respectively.

By comparing the signals received from a pair of reference and surveillance channels, the Doppler frequencies raised by the mobile blocker can be detected at a particular AoA. Due to the carrier frequency offset (CFO) and the sampling clock frequency offset (SFO) between the transmitter and the receiver, the time-of-fly (ToF) measurement of reference channels or surveillance channels could be difficult. It is therefore infeasible to localize the mobile blocker at the receiver via single capture of the blocker. Instead, the Doppler frequency and AoA of the mobile blocker are successively tracked in the proposed method, so that the trajectory of the mobile blocker as well as the locations of the two transmitters can be jointly detected with a sufficient number of observations.

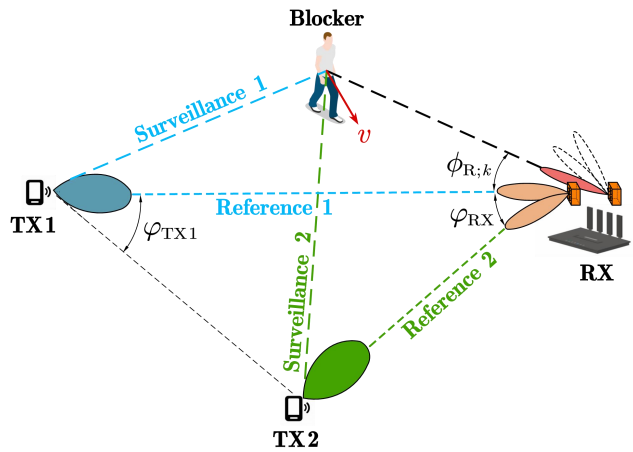

Figure 1. Overview of the integrated sensing and communication system

## 3 Signal Processing of Cooperative Passive Sensing

**3.1 Signal Model**

It is assumed that transmission and receiving beams of the two reference channels have been aligned via the existing method, e.g., exhaustive beam search. On the other hand, the surveillance beam switches sequentially and periodically among Q-directions, denoted as $\Phi = \{\phi_1, \phi_2, \cdots, \phi_Q\}$. In each direction, signals of both frequency bands are received via the surveillance beam for a duration of $T_b$ s. Thus, it takes the surveillance beam $T_d = QT_b$ s to complete a sweeping period.

In the $k$-th sweeping period ($\forall k$), let $s_{m,k,q}(t)$ be the transmit baseband signal of $m$-th transmitter ($m = 1,2$) when the surveillance beam is at the direction $\phi_q$, and the received signal via the $m$-th reference beam in the $m$-th frequency band can be written as:

$$y_{m,k,q}^{\text{ref}}(t) = \alpha_{m,k,q}^{\text{ref}} s_{m,k,q}\left(t - \tau_{m,k,q}^{\text{ref}}\right) + n_{m,k,q}^{\text{ref}}(t), 0 \leq t \leq T_b, \tag{1}$$

where $\alpha_{m,k,q}^{\text{ref}}$ and $\tau_{m,k,q}^{\text{ref}}$ denote the complex gain and delay of the LoS path, $n_{m,k,q}^{\text{ref}}(t)$ denotes the superposition of noise and NLoS echoes. As a remark, the scattered signal of $s_{m,k,q}(t)$ may also be received by the reference beam.

Simultaneously, the received signal of the surveillance beam in the $m$-th frequency band, denoted as $y_{m,k,q}^{\text{sur}}(t)$, includes the scattered signals from the mobile blocker and static scattering clusters. Thus,

$$y_{m,k,q}^{\text{sur}}(t) = \alpha_{m,k,q}^{\text{tar}}(t) s_{m,k,q}\left(t - \tau_{m,k,q}^{\text{tar}}(t)\right) e^{-j2\pi f_{m,k,q}^{\text{tar}}(t)t} + \sum_{l=1}^{L_{m,q}} \alpha_{m,k,q}^{l} s_{m,k,q}\left(t - \tau_{m,k,q}^{l}\right) + n_{m,k,q}^{\text{sur}}(t), 0 \leq t \leq T_b, \tag{2}$$

where $\alpha_{m,k,q}^{\text{tar}}(t)$, $\tau_{m,k,q}^{\text{tar}}(t)$ and $f_{m,k,q}^{\text{tar}}(t)$ denote the time-varying complex gain, delay and Doppler frequency of the surveillance channel, respectively, $L_{m,q}$ denotes the number of NLoS paths via static scattering clusters, $\alpha_{m,k,q}^{l}$ and $\tau_{m,k,q}^{l}$ are the complex gain and delay of the $l$-th one, and $n_{m,k,q}^{\text{sur}}(t)$ denotes the noise. As a remark, the signal from LoS path may also be received by the surveillance beam, which can be treated as a special static scattering cluster in the second term of the above equation.

The received signals from both reference and surveillance beams in the two frequency bands are sampled with a period $T_s$, which can be expressed by $y_{m,k,q}^{\text{ref}}[n] = y_{m,k,q}^{\text{ref}}(nT_s)$ and $y_{m,k,q}^{\text{sur}}[n] = y_{m,k,q}^{\text{sur}}(nT_s)$, where $n = 1,2,\cdots,T_b/T_s$, $m = 1,2$, $q = 1,2,\ldots,Q$. Note that the signal components in $y_{m,k,q}^{\text{sur}}[n]$ with non-zero Doppler frequencies from the moving blocker may be overwhelmed by the strong scattered signals with zero Doppler frequency. Specifically, in Eq. (2), the first term on the right side is far smaller than the second term. This can disrupt the estimation of Doppler frequency. Hence, the least-square-based (LS-based) clutter cancellation in Ref. [17] is applied to suppress the signal components with zero Doppler frequency in $y_{m,k,q}^{\text{sur}}[n]$. The surveillance signal after clutter cancellation is denoted as $\hat{y}_{m,k,q}^{\text{sur}}[n]$.

### 3.2 Doppler Frequency and AoA Estimation

The Doppler frequency estimation in passive sensing is based on the cross-ambiguity function (CAF) between the reference signals and surveillance signals. Particularly, the CAF of the received signals in the $m$-th frequency band, $k$-th sweeping period, and $q$-th surveillance beam's direction is defined as

$$R_{m,k}(q, f_d) = \max_{\tau_{m,k,q}} \sum_{n=1}^{N_0} \hat{y}_{m,k,q}^{\text{sur}}[n] \left\{ y_{m,k,q}^{\text{ref}}[n - \tau_{m,k,q}] \right\}^* e^{j2\pi f_d nT_s}, \tag{3}$$

where $\{.\}^*$ is the complex conjugate; $N_0 = T_b/T_s$ denotes the number of samples when the surveillance beam is in one direction. Since we only focus on the estimation of the Doppler frequency, the delay $\tau_{m,k,q}$ is not considered as a parameter of the CAF. There should be a peak value of $R_{m,k}(q, f_d)$ at $f_d = f_{m,k,q}^{\text{tar}}(kN_0 T_s)$. Thus, Doppler frequencies of the mobile blocker could be detected by finding the peak values of $R_{m,k}(q, f_d)$. As a remark, note that the estimation of the Doppler frequency in Eq. (3) does not request a priori knowledge on the path gains $\alpha_{m,k,q}^{\text{tar}}$ and $\alpha_{m,k,q}^{l}$.

There might be more than one peak value of $R_{m,k}(q, f_d)$ in the $k$-th sweeping period. This is because there might be multiple scattering points on the blocker with different velocities. The scattering between the blocker and the surrounding clutters would also generate weak peaks on $R_{m,k}(q, f_d)$. An adaptive-threshold-based method is adopted to detect the Doppler frequencies with the dominant receiving power from the CAF. First, the threshold for the Doppler frequency $f_d$ in the $q$-th surveillance beam's direction can be calculated as:

$$T_{m,k}(q, f_d) = \frac{\gamma}{2W+1} \sum_{p=-W_T}^{W_T} R_{m,k}(q, f_d + p\Delta f), \tag{4}$$

where $W_T$ is the half length of training cells, $\gamma > 1$ is a scaling factor for the detection threshold, and $\Delta f = 1/(N_0 T_s)$ is the resolution of the Doppler frequency. Thus, in the $k$-th sweeping period, a Doppler frequency $f_d$ is detected in the $q$-th surveillance beam's direction when $R_{m,k}(q, f_d) \geq T_{m,k}(q, f_d)$.

Note that the scattered signal from the mobile blocker might be captured by multiple beam directions, leading to the false alarm in the AoA detection. In the proposed system, we treated the surveillance beam's direction maximizing the CAF as the estimated AoA of each sweeping period. Particularly, denoting $\tilde{f}_{m,k}$ and $\tilde{\phi}_k$ as the estimated Doppler frequency and AoA in the $k$-th sweeping period, we have

$$(\tilde{f}_{m,k}, \tilde{q}_{m,k}) = \underset{f_d,\, q}{\arg\max} R_{m,k}(q, f_d) \quad\quad\quad (5)$$
$$\text{s.t.} \quad R_{m,k}(q, f_d) \geq T_{m,k}(q, f_d)$$

and $\tilde{\phi}_k = \phi_{\tilde{q}_{m,k}}$, respectively. Note that the AoA of blockers is independent of the frequency bands. Finally, we define the measured feature vector $\mathbf{z}_k$ of the $k$-th sweeping periods as

$$\mathbf{z}_k = [\tilde{\phi}_k, \tilde{f}_{1,k}, \tilde{f}_{2,k}]^\text{T}. \quad\quad\quad (6)$$

## 4 Localization and Blockage Prediction

In this section, a joint estimation method is proposed to detect the positions of the two transmitters and the trajectory of the mobile blocker based on the measured AoAs and Doppler frequencies in a number of sweeping periods. The geometric relation among the two transmitters, the receiver and the mobile blocker is illustrated in Fig. 2. Without loss of generality, the coordinates of the receiver and the two transmitters are represented by vectors $\mathbf{p}^\text{RX} = [0,0]^\text{T}$, $\mathbf{p}_1^\text{TX} = [d, 0]^\text{T}$, $\mathbf{p}_2^\text{TX} = [x_\text{TX2}, y_\text{TX2}]^\text{T}$, respectively, and the coordinates of the mobile blocker in the $k$-th sweeping period are represented by $\mathbf{p}_k = [x_k, y_k]^\text{T}$. There is no a priori knowledge on the locations of the two transmitters at the receiver. Thus, $d$, $x_\text{TX2}$ and $y_\text{TX2}$ are unknown.

To facilitate the joint estimation, it is assumed that the two transmitters and a receiver have sent and received signals to and from each other, so that the angles-difference-of-arrival (ADoAs) between them have been estimated via the spatial smoothing multiple signal classification (MUSIC) algorithm[18]. Thus, the angles $\varphi_\text{RX}$ and $\varphi_\text{TX1}$ in Fig. 2 are known at the receiver. Note that $x_\text{TX2}$ and $y_\text{TX2}$ are functions of $d$, $\varphi_\text{RX}$ and $\varphi_\text{TX1}$, the joint estimation of $d$ and $\mathbf{p}_k$, $k = 1,2,3,...$, is elaborated in this section, followed by the blockage prediction. In the following, we first elaborate the motion model of the mobile blocker.

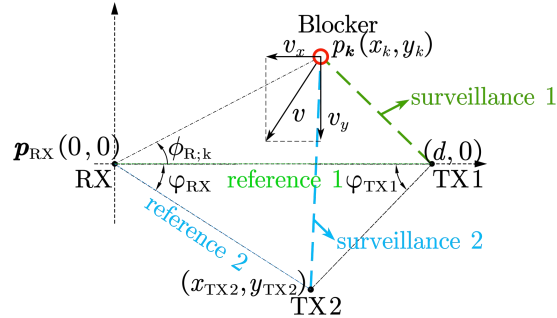

Figure 2. Illustration of the blocker tracking system in the coordinate system with two transmitters.

**4.1 Motion Model**

According to the geometric relation, the ADoA $\varphi_\text{RX}$ and $\varphi_\text{TX1}$ can be expressed in terms of the coordinates of the two transmitters and the receiver as:

$$\begin{cases} \varphi_\text{RX} = \cos^{-1} \frac{(\mathbf{p}_1^\text{TX}-\mathbf{p}^\text{RX})^\text{T}(\mathbf{p}_2^\text{TX}-\mathbf{p}^\text{RX})}{\|\mathbf{p}_1^\text{TX}-\mathbf{p}^\text{RX}\|\|\mathbf{p}_2^\text{TX}-\mathbf{p}^\text{RX}\|} \\ \varphi_\text{TX1} = \cos^{-1} \frac{(\mathbf{p}^\text{RX}-\mathbf{p}_1^\text{TX})^\text{T}(\mathbf{p}_2^\text{TX}-\mathbf{p}_1^\text{TX})}{\|\mathbf{p}^\text{RX}-\mathbf{p}_1^\text{TX}\|\|\mathbf{p}_2^\text{TX}-\mathbf{p}_1^\text{TX}\|} \end{cases}. \quad\quad\quad (7)$$

Since the sweeping period is very short, it can be approximated that the mobile blocker is moving with a constant velocity in a sweeping period. Hence, the trajectory of the mobile block can be expressed as:

$$\mathbf{p}_k = \mathbf{p}_{k-1} + \mathbf{v}_{k-1} T_d = \mathbf{p}_1 + \sum_{n=1}^{k-1} \mathbf{v}_n T_d, \quad\quad\quad (8)$$

where $\mathbf{v}_k = [v_{\text{x};k}, v_{\text{y};k}]^\text{T}$ is the vector of velocity in the $k$-th sweeping period; $v_{\text{x};k}$ and $v_{\text{y};k}$ denote the velocity components in the $x$-axis and $y$-axis, respectively;

Hence, the AoA of the surveillance channel in the $k$-th sweeping period $\phi_{R;k}$ can be written as

$$\phi_{R;k} = \cos^{-1}\frac{p_k}{\|p_k\|} = \cos^{-1}\frac{p_1+\sum_{n=1}^{k-1}v_n T_d}{\|p_1+\sum_{n=1}^{k-1}v_n T_d\|}. \tag{9}$$

Moreover, the Doppler frequency of the $m$-th surveillance channel is given by

$$f_{d;m,k} = \frac{1}{\lambda_m}\left(\frac{p_k-p_m^{TX}}{\|p_k-p_m^{TX}\|}+\frac{p_k-p^{RX}}{\|p_k-p^{RX}\|}\right)^T v_k, \tag{10}$$

where $\lambda_m$ is the carrier frequency of the $m$-th transmitter.

### 4.2 Joint Localization and Trajectory Estimation

It can be observed from the above motion model that, given the distance $d$, the initial position $p_1$ and the velocity vectors $v_1, v_2, ..., v_k$ of the blocker, the AoAs and Doppler frequencies of all the sweeping periods can be calculated according to Eq. (9) and Eq. (10). Let $x_k = [d, p_1^T, v_1^T, v_2^T, \cdots, v_k^T]^T$ be the vector of all the motion parameters to be estimated in the first $k$ sweeping periods; $h_k = [\phi_{R;k}, f_{d;1,k}, f_{d;2,k}]^T$ is the true feature vector of $k$-th sweeping period with motion parameters $x_k$; $H(x_k) = [h(x_1), h(x_2), \cdots, h(x_k)]^T$ and $Z_k = [z_1, z_2, \cdots, z_k]^T$ are the aggregation of the true feature vectors and measured feature vectors, respectively. The trajectory of the mobile blocker and the positions of transmitters can be estimated via minimizing the difference between true features and the measured ones in the first $K$ sweeping periods. Thus,

$$\widehat{x}_K = \underset{x_K}{\arg\min}\, tr\{W[Z_K - H(x_K)]^T[Z_K - H(x_K)]\}, \tag{11}$$

where $tr\{.\}$ denotes the matrix trace. $W = \mathrm{diag}(\alpha_1, \alpha_2, \alpha_3)$ is the weighting matrix, and $\alpha_1, \alpha_2, \alpha_3$ denote the weights of different features, respectively.

Although the above problem is nonlinear, we can find local optimal solutions by the classical Levenberg-Marquardt (LM) optimization algorithm[19–20] with multiple initial solutions. Particularly, we first define function $f(x_K) = tr\{W[Z_K - H(x_K)]^T[Z_K - H(x_K)]\}$, and let $x_{K,1}^0, x_{K,2}^0, \cdots, x_{K,S}^0$ be the $S$ initial estimations of the motion parameters $x_K$. In the $l$-th iteration ($l = 1,2,3,\cdots$), the estimations of motion parameters $x_K$ are updated as $x_{K,s}^l = x_{K,s}^{l-1} + J(x_{K,s}^{l-1})\Delta$, $s = 1,2,3,\cdots,S$, where $J(x_K) = \frac{\partial f(x_K)}{\partial x_K}$ is the Jacobian matrix and $\Delta$ is the step size. Finally, let $L$ be the total number of iterations, and the estimated motion parameters are given by

$$\widehat{x}_K = \underset{x_K \in \{x_{K,1}^L, x_{K,2}^L, \cdots, x_{K,S}^L\}}{\arg\min}\, f(x_K). \tag{12}$$

### 4.3 Blockage Detection

Based on the estimated trajectory and velocities of the blocker in the $K$ sweeping periods, it is feasible to predict whether and when the LoS between transmitters and the receiver would be blocked. Assuming that the mobile blocker keeps the average velocity of the previous $n$ sweeping periods, the predicted trajectory $p(t)$ of the blocker after the $K$-th sweeping period can be expressed as

$$p(t) = \widehat{p}_K + \overline{v}t, \tag{13}$$

where $\overline{v} = \sum_{k=K-n+1}^{K} \widehat{v}_k$ is the average velocity in previous $n$ sweep periods, and $\widehat{v}_k$ is the estimated velocity of the blocker in the $k$-th sweeping period according to Eq. (12).

Let $b_m \in \{0,1\}$, $m = 1,2$, be the blockage indicator, where $b_m = 1$ denotes that the $m$-th LoS will be blocked in the future and $b_m = 0$ otherwise. We have

$$b_m = \begin{cases} 1, & \det([\overline{v}, p^{RX} - \widehat{p}_K])\det([\overline{v}, p_m^{TX} - \widehat{p}_K]) < 0 \\ 0, & \text{otherwise} \end{cases}. \tag{14}$$

Moreover, let $\hat{t}_m$ be the estimated blockage time, after which the link blockage will happen. Let $\mu p_m^{TX}$ be the position of the intersection point in the $m$-th LoS path, where $\mu \in [0,1]$. The blockage time $\hat{t}_m$ and $\mu$ can be calculated by

$$\begin{bmatrix}\mu \\ \hat{t}_m\end{bmatrix} = [p_m^{TX}, -\overline{v}]^{-1}\widehat{p}_K, \tag{15}$$

where $\det([\boldsymbol{p}_m^{\text{TX}}, -\overline{\boldsymbol{v}}]) \neq 0$.

## 5 Experiment Results and Analysis

In the experiments, the implementation of the proposed system is shown in Fig. 3. Both Transmitters 1 and 2 are implemented with one NI USRP-2954R connected with one Sivers 60 GHz phased array. In both transmitters, the transmitting signal with a bandwidth of 1 MHz consists of a training sequence and an orthogonal frequency division multiplexing (OFDM)-modulated data payload. The carrier frequencies of Transmitters 1 and 2 are 60.98 GHz and 60.985 GHz, and the width of transmitting beams are both 90°. At the receiver, two Sivers 60 GHz phased arrays, with a clock of 45 MHz, are connected to one NI USRP-2954R. One of the phased arrays adopts a two-lobe receiving beam towards Transmitters 1 and 2, respectively, which receives signals of reference channel 1 and 2 in two different frequency bands. The surveillance beam is implemented via the other phased array, covering both frequency bands. Signals from two transmitters can be separated at the USRP of the receiver via two bandpass filters. Moreover, the surveillance beam is switched periodically among $Q=4$ directions, which are 40°, 27°, 18° and 10°, with a beamwidth of about 10°. The sensing duration of one direction is $T_b$=50 ms and the duration of one sweeping period is $T_d = 200$ ms which implies the minimum detectable speed is around 0.025 m/s. The sampling frequency at the baseband is 10 MHz.

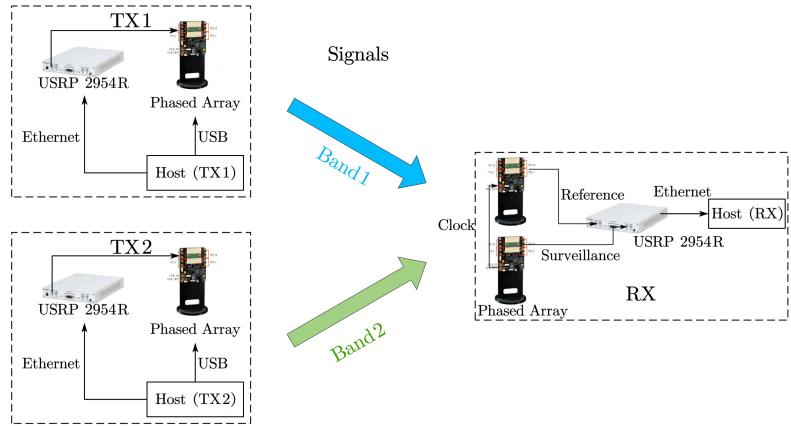

Figure 3. Block diagram of system implementation

The experiments are conducted in a laboratory environment with rich scattering clusters, e.g., displays and metallic cabinets. The placement of the two transmitters and the receiver platform is illustrated in Fig. 4, where all of the phased arrays are placed at a height of 1.5 m. Taking the receiver as the origin, the coordinates of Transmitters 1 and 2 are $\boldsymbol{p}_1^{\text{TX}} = [2.7\ m, 0]^{\text{T}}$ and $\boldsymbol{p}_2^{\text{TX}} = [1.8\ m, -1.4\ m]^{\text{T}}$. A volunteer is walking in this region as the mobile blocker. A depth camera system (ZED) is deployed behind the receiver to record the true trajectory of the blocker. It is synchronized with the receiver at the millisecond level. Since different parts of the body have different trajectories, we extract the trajectory information of 21 key points of the human body from the depth camera[21] with centimeter-level accuracy, and represent the real trajectory of the human body by that of the "neck" keypoint. This is because the keypoint of "neck" is at the same height as the phased array.

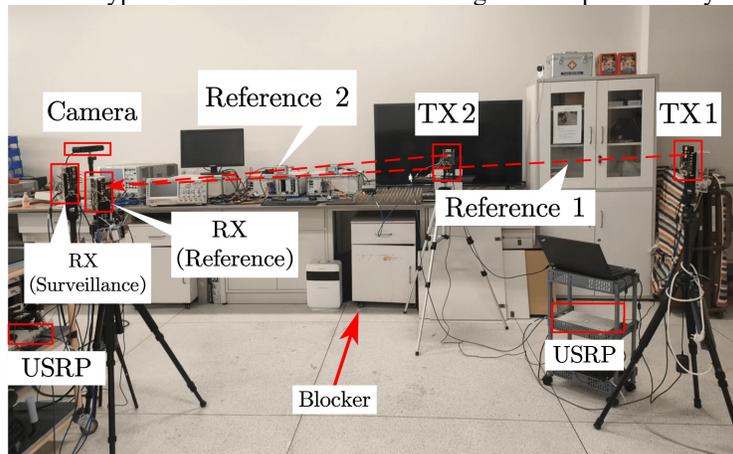

USRP: Universal Software Radio Peripheral
Figure 4. Illustration of the experiment environment

## 5.1 Detection of Doppler Frequency and AoA

As an example, the Doppler-angle spectrograms of two surveillance channels in one sweeping period are shown in Fig. 5. In both spectrograms, the strengths of the signal components versus beam directions (AoAs) and Doppler frequencies are illustrated by colors. It can be observed that a Doppler frequency of 140 Hz is detected in the surveillance channel 1, and a Doppler frequency of 300 Hz is detected in the surveillance 2. Thus, the same movement of the blocker would generate different Doppler frequencies at the two surveillance channels due to different locations of the transmitters. It can also be observed that there are multiple peaks of Doppler frequencies in Fig. 5 (b) that are caused by different movements of scattering points on the human body, e.g., 350 Hz, 300 Hz and 270 Hz. According to Eq. (5), we will choose the Doppler frequency with the highest signal strength as the feature of the sweeping period. As a result, since the Doppler frequencies are mainly captured by the second beam direction, the AoA of the sweeping period is 27°.

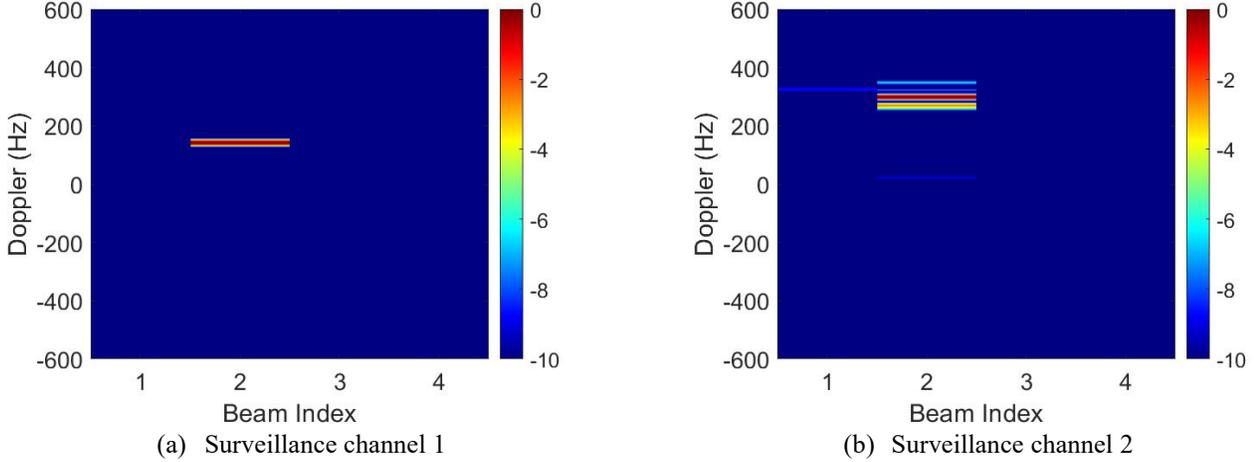

(a) Surveillance channel 1          (b) Surveillance channel 2

Figure 5. Doppler frequency of 4 beams (beams 1 to 4 indicate AoAs of 40°, 27°, 18° and 10°) in a sweeping period

All detected Doppler frequencies and AoAs versus time of a trajectory are shown in Fig. 6. It can be observed that there is no moving target in the time interval [0, 1.2 s], since no significant Doppler frequency can be detected. The Doppler frequencies of surveillance channel 1 is consistently greater than that of surveillance channel 2, and the Doppler frequency of the first surveillance channel is close to 0 in [2.8 s, 3.2 s]. This is because the projection of velocity along the surveillance channel 2 is more significant according to locations of both transmitters. Furthermore, since the estimation error of AoAs could be large, it is necessary to smooth AoAs via polynomials as shown by the blue line. After the detection of Doppler frequencies and AoAs of the two surveillance channels, the trajectory of the blocker and the positions of transmitters can be obtained by solving the problem in Eq. (12).

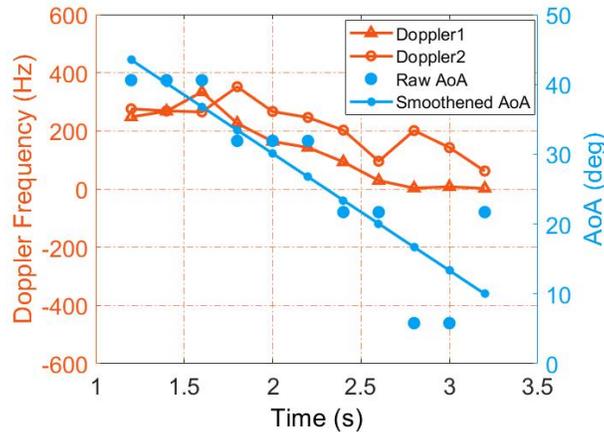

Figure 6. Estimated Doppler frequencies of two surveillance channels and AoAs

## 5.2 Trajectory Estimation and Transmitter's Localization

In the experiments, it is estimated that the ADoAs are $\varphi_{RX} = 39°$ and $\varphi_{RX} = 58°$. Hence, the locations of the two transmitters and the trajectory of the mobile blocker can be estimated jointly according to the above estimated Doppler frequencies and AoAs. The trajectories of the two trails are illustrated in Figs. 7(a) and 7(b), respectively, where the red curve is the true trajectory of the blocker recorded by the ZED and the green one is the estimated one. Moreover, the true and estimated locations of the two transmitters are also differentiated by colors. It can be observed that the localization errors of the two transmitters are within 0.2 m. The estimated trajectory is almost parallel to the real trajectory, but with an offset in *x*-axis. The average error in the blocker's trajectory and raw AoA estimation is 0.39 m and 7.8 degrees, respectively, with corresponding mean squared errors of $0.48^2$ and $9.4^2$. Nevertheless, the estimated trajectory is sufficiently accurate in the blockage prediction.

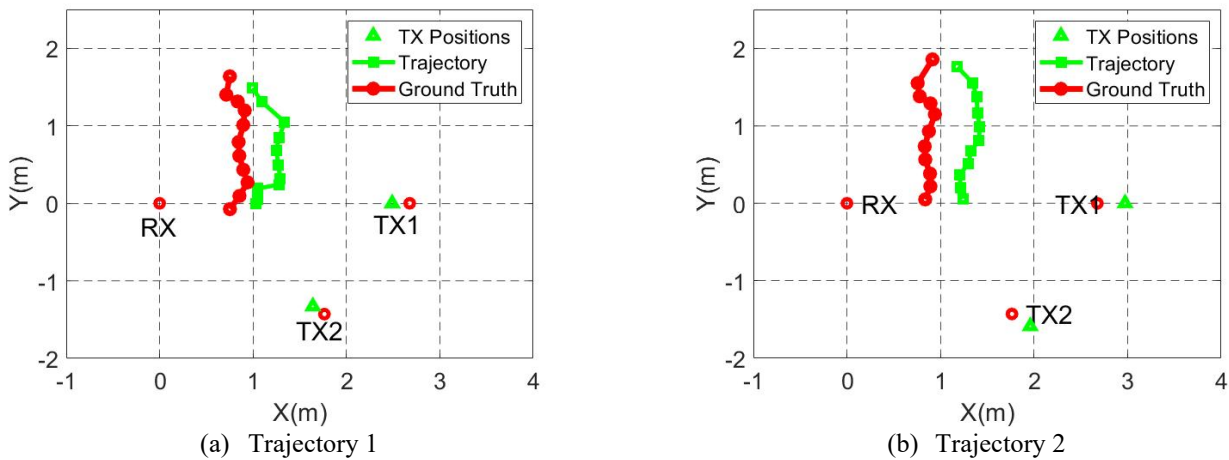

(a) Trajectory 1    (b) Trajectory 2

Figure 7. Estimated trajectory of the blocker and the positions of transmitters

**5.3 Blockage Prediction**

The prediction of blockage time, which measures the remaining time duration until link blockage, for the above two trajectories are illustrated in Fig. 8. Due to the placement of transmitters and the receiver, the link blockage will happen in reference channel 1. It can be observed that, the error of blockage time $\hat{t}_1$ is less than 0.1 s when the blockage time is less than 0.6 s. Moreover, larger blockage time leads to larger prediction error. This is due to the time-varying velocity of the volunteer.

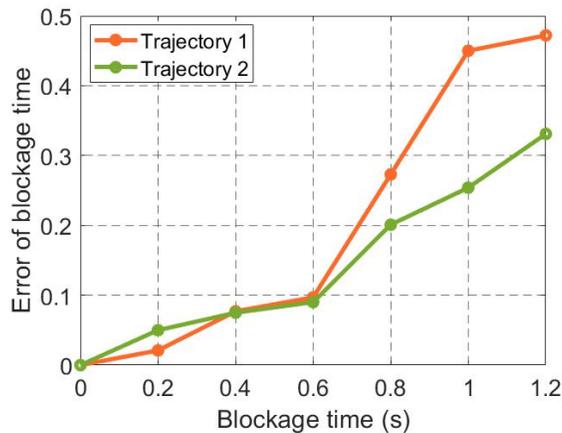

Figure 8. Error of predicted blockage time $\hat{t}_1$

# 6 Conclusions

In the paper, a method that employs passive sensing techniques to jointly estimate transmitter positions and blocker trajectories is proposed for mmWave communication systems, so that the link blockage and the blockage time can be predicted. The method can be deployed in the uplink transmission scenario with one mobile blocker and at least two transmitters. Without a priori knowledge on the positions of transmitters, the trajectory of the mobile blocker is estimated by tracking its Doppler frequencies and AoAs in a number of sweeping periods, where a gradient-descent algorithm is proposed to suppress the estimation error. It is demonstrated that the system can achieve decimeter-level localization and trajectory estimation, and the blockage time prediction error is within 0.1 s, when the blockage time is less than 0.6 s.

While this paper considers the tracking of a single blocker, the proposed system has the potential to identify and track multiple blockers with distinct Doppler frequencies or AoAs. Moreover, it is possible to improve the detection accuracy of the distances between the transmitters and the receiver by observing a number of trajectories of a single blocker in different trials. Hence, the detection accuracy of a single blocker's trajectory could also be further improved. We would like to investigate the above extensions as our future research.